\documentclass{article}

\usepackage{makeidx}
\usepackage{amsthm,amsmath,amssymb,amsfonts}
\usepackage{setspace,graphicx}
\usepackage{Generic}  
\usepackage[sort, longnamesfirst]{natbib}
\usepackage{url} 

\usepackage[usenames,dvipsnames,table]{xcolor}
\usepackage{tikz}

\usepackage{changepage}

\usepackage{hyperref}
\hypersetup{
    colorlinks=true,
    linkcolor=red,
    citecolor=MidnightBlue,
}

\definecolor{charlesBlue}{RGB}{100, 155, 255}
\definecolor{lightgray}{RGB}{200, 200, 200}

\numberwithin{equation}{section}
\theoremstyle{plain}

\usepackage{setspace}

\begin{document}
\onehalfspacing
%



\title{\bf{For how many iterations should we run \\ Markov chain Monte Carlo?}}
\author{Charles C. Margossian\\
{\small Center for Computational Mathematics, Flatiron Institute, New York, NY} \\[0.1in]
Andrew Gelman\\
{\small Department of Statistics and Political Science, Columbia University, New York, NY}
}
\date{}

\maketitle

\begin{flushright}
``All human wisdom is contained in \\ these two words: wait and hope.'' \\[0.1in]
 --- Alexandre Dumas,\\ 
\textit{The Count of Monte Cristo}
\end{flushright}

\begin{abstract}

Standard Markov chain Monte Carlo (MCMC) admits three fundamental control parameters: the number of chains, the length of the warmup phase, and the length of the sampling phase.
These control parameters play a large role in determining the amount of computation we deploy.
In practice, we need to walk a line between achieving sufficient precision and not wasting precious computational resources and time.
We review general strategies to check the length of the warmup and sampling phases, and examine the three control parameters of MCMC in the contexts of CPU- and GPU-based hardware.
Our discussion centers around three tasks: (1) {\bf inference} about a latent variable, (2) {\bf computation} of expectation values and quantiles, and (3) {\bf diagnostics} to check the reliability of the estimators.

This chapter begins with general recommendations on the control parameters of MCMC, which have been battle-tested over the years
and often motivate defaults in Bayesian statistical software.
Usually we do not know ahead of time how a sampler will interact with a target distribution,
and so the choice of MCMC algorithm and its control parameters, tend to be based on experience, re-evaluated after simulations have been obtained and analyzed.
The second part of this chapter provides a theoretical motivation for our recommended approach, with pointers to some concerns and open problems.
We also examine recent developments on the algorithmic and hardware fronts, which motivate new computational approaches to MCMC.

\end{abstract}

\section*{Part I: Practical considerations}

If well constructed, an MCMC sampler converges to a stationary distribution $p$ of interest and, given enough computation time, the resulting Monte Carlo estimators can achieve any precision with high probability.
In practice the asymptotic properties of MCMC at best approximate the behavior of MCMC under a finite computational budget.
Hence we must carefully reason about the behavior of finite nonstationary Markov chains and understand how approximate samples---ultimately not drawn from the stationary distribution $p$---can still produce useful Monte Carlo estimators.
Approximate convergence is a means of controlling the \textit{bias} of our estimators, and so our goal is not to generate samples from $p$, rather from an approximation $\hat p$ that incurs an acceptable bias.
We must then generate enough samples to reduce the \textit{variance} of our estimators.
These objectives respectively inform the warmup and the sampling phases of MCMC.  


\section{Recommendations for an MCMC workflow} \label{sec:recommendations}

\begin{figure}
\begin{center}
\resizebox{4.75in}{1.5in}{%
\begin{tikzpicture}
    [
       Orange/.style={circle, draw=orange!, fill=green!0, thick, minimum size=10},
       Blue/.style={circle, draw=charlesBlue!, fill=green!0, thick, minimum size=10},
       Empty/.style={, draw=white!, fill=green!0, minimum size=0mm}
    ]

    \node[Empty] (p0) at (0, 0.5) {$p_0$};
    \node[Empty] (phat) at (5, 0.5) {$\hat p \approx p$};

    \node[Empty] (time) at(2, -1.5) {$\underbrace{\hspace{4.5cm}}$};
    \node[Empty] (warmup) at (2, -2) {\small warmup phase};

     \node[Empty] (time) at(7, -1.5) {$\underbrace{\hspace{4.5cm}}$};
    \node[Empty] (warmup) at (7, -2) {\small sampling phase};
    
    \node[Orange] (theta0) at (0, 0) {};
    \node[Orange] (theta1) at (1, 0) {};
    \node[Orange] (theta2) at (2, 0) {};
    \node[Orange] (theta3) at (3, 0) {};
    \node[Orange] (theta4) at (4, 0) {};
    \node[Blue] (theta5) at (5, 0) {};
    \node[Blue] (theta6) at (6, 0) {};
    \node[Blue] (theta7) at (7, 0) {};
    \node[Blue] (theta8) at (8, 0) {};
    \node[Blue] (theta9) at (9, 0) {};

    \node[Orange] (1theta0) at (0, -0.5) {};
    \node[Orange] (1theta1) at (1, -0.5) {};
    \node[Orange] (1theta2) at (2, -0.5) {};
    \node[Orange] (1theta3) at (3, -0.5) {};
    \node[Orange] (1theta4) at (4, -0.5) {};
    \node[Blue] (1theta5) at (5, -0.5) {};
    \node[Blue] (1theta5) at (5, -0.5) {};
    \node[Blue] (1theta6) at (6, -0.5) {};
    \node[Blue] (1theta7) at (7, -0.5) {};
    \node[Blue] (1theta8) at (8, -0.5) {};
    \node[Blue] (1theta9) at (9, -0.5) {};

    \node[Orange] (2theta0) at (0, -1) {};
    \node[Orange] (2theta1) at (1, -1) {};
    \node[Orange] (2theta2) at (2, -1) {};
    \node[Orange] (2theta3) at (3, -1) {};
    \node[Orange] (2theta4) at (4, -1) {};
    \node[Blue] (2theta5) at (5, -1) {};
    \node[Blue] (2theta6) at (6, -1) {};
    \node[Blue] (2theta7) at (7, -1) {};
    \node[Blue] (2theta8) at (8, -1) {};
    \node[Blue] (2theta9) at (9, -1) {};

    \draw [-latex](theta0) -- (theta1);
    \draw [-latex](theta1) -- (theta2);
    \draw [-latex](theta2) -- (theta3);
    \draw [-latex](theta3) -- (theta4);
    \draw [-latex](theta4) -- (theta5);
    \draw [-latex](theta5) -- (theta6);
    \draw [-latex](theta6) -- (theta7);
    \draw [-latex](theta7) -- (theta8);
    \draw [-latex](theta8) -- (theta9);

    \draw [-latex](1theta0) -- (1theta1);
    \draw [-latex](1theta1) -- (1theta2);
    \draw [-latex](1theta2) -- (1theta3);
    \draw [-latex](1theta3) -- (1theta4);
    \draw [-latex](1theta4) -- (1theta5);
    \draw [-latex](1theta5) -- (1theta6);
    \draw [-latex](1theta6) -- (1theta7);
    \draw [-latex](1theta7) -- (1theta8);
    \draw [-latex](1theta8) -- (1theta9);
    
    \draw [-latex](2theta0) -- (2theta1);
    \draw [-latex](2theta1) -- (2theta2);
    \draw [-latex](2theta2) -- (2theta3);
    \draw [-latex](2theta3) -- (2theta4);
    \draw [-latex](2theta4) -- (2theta5);
    \draw [-latex](2theta5) -- (2theta6);
    \draw [-latex](2theta6) -- (2theta7);
    \draw [-latex](2theta7) -- (2theta8);
    \draw [-latex](2theta8) -- (2theta9);

    \end{tikzpicture}
    }
\end{center}
\caption{\textit{Standard MCMC workflow. We initialize at least three Markov chains from a starting distribution $p_0$. 
We then run a warmup phase, during which the Markov chains move from $p_0$ to a warm distribution $\hat p$, close to the target distribution $p$.
The warmup is primarily used to reduce bias and tune the sampler.
We then discard the warmup iterations and run a sampling phase, during which we collect samples to reduce variance.
}}
\label{fig:chains}
\end{figure}
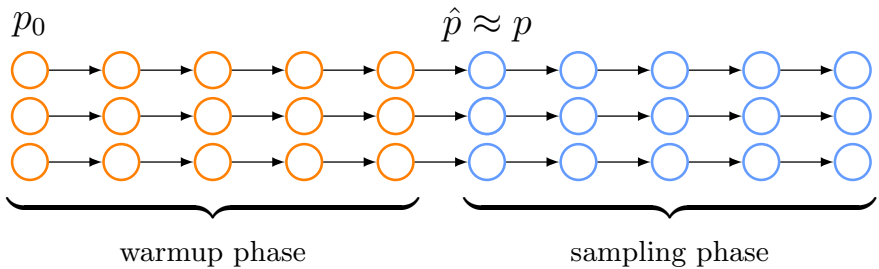

We begin by reviewing a typical MCMC workflow (Figure~\ref{fig:chains}).
Many of these recommendations are implemented in state-of-the-art statistical software for Bayesian analysis, including \texttt{Stan} \citep{Carpenter:2017}, \texttt{PyMC} \citep{Salvatier:2016}, and others, and have been applied across a range of statistical problems.

\begin{enumerate}
    \item Simulate three or more chains with distinct initializations, for instance based on a crude estimate of the target distribution $p$.
    In a Bayesian context, we may elect to use the prior on the model parameters (if these are proper) to generate starting points.
    We can typically run 4 or 8 chains in parallel on a modern laptop and more when using a computer cluster. 
    Having multiple chains with different initializations and seed can help diagnose issues with convergence, and reduces the variance of our Monte Carlo estimator.
    
    \item Each Markov chain should be split into a warmup phase, during which we allow the MCMC sampler to learn its tuning parameter \citep{Andrieu:2008}, and a sampling phase during which we freeze the tuning parameters.
    Only use the sampling iterations to construct Monte Carlo estimates.
    The warmup iterations may be saved to diagnose issues with our Markov chains.
    As a starting point, we recommend splitting the chains roughly half-way into a warmup and a sampling phase.
    The warmup phase allows us to learn the tuning parameter of the sampler and reduce the bias due to an imperfect initialization; the sampling phase's primary role, on the other hand, is to reduce the variance of our estimators.
    
    \item Check that, despite the distinct seeds and initializations, the Markov chains all produce estimates which are in reasonable agreement.
    This can be assessed using visual tools, such as trace and density plots, or convergence diagnostics such as $\widehat R$ \citep{Gelman:1992, Vehtari:2021}.
    In addition, it can be useful to split each chain into subchains to identify transient behaviors \citep{Gelman:2013}.
    Some samplers, such as Hamiltonian Monte Carlo (HMC) \citep{Neal:2011, Betancourt:2018}, provide additional diagnostics such as divergent transitions, which can indicate poor mixing.
    These diagnostics are complementary, in the sense that they may identify failures even when $\widehat R$ looks fine.
    In general, it is a good idea to use multiple diagnostics, since no diagnostic alone is completely reliable \citep{Brooks:2003}.
    
    \item If approximate convergence is diagnosed, examine the variance of the Monte Carlo estimators using the Monte Carlo standard error (MCSE) and the effective sample size (ESS) for any quantity of interest.
    Be mindful that the ESS is specific to a Monte Carlo estimator and so can vary between two estimators, even when they are computed using samples from the same Markov chains.
    You should have a rough understanding of the precision required for your problem.
    Targeting a number of significant digits can be expressed as a target ESS \citep{Vats:2019, Vehtari:2022}.

    \item For a complicated model, consider first fitting a simpler model and building up.
    The simpler models can serve as a baseline.
    A short run of MCMC can also help us spot obvious errors and get a better sense of how much computation may be required.


    
\end{enumerate}

A difficult question is what to do when the Markov chains fail to converge (reduce bias) or mix (reduce variance).
Throwing more computation at the problem---either by running more iterations, changing the tuning parameters of the sampler, or the sampler altogether---may help but it is not always the best approach.
In our experience, computational problems can indicate modeling problems (the ``folk theorem of computational statistics'') and we may try choosing better starting points, debugging the model, reparameterizing, improving identification (simplifying or changing the model or adding prior information if available), etc; see \citet{Gelman:2020} for a review.

Beyond that, since the development of \texttt{Stan} and more broadly popular Bayesian inference software based on HMC \citep{Strumbelj:2023}, there have been several methodical and hardware developments which can motivate departures from the current defaults.

\section{Inferential goals}

Sound use of computation requires understanding how much precision is useful at various stages of an analysis and estimating this precision.
Following \citet{Gelman:2011}, we distinguish two sorts of inference or computation tasks:
\begin{itemize}

  \item[] {\bf Task 1. Inference} about an unknown quantity $\theta$, or more generally any quantity of interest $f(\theta)$.
  Such inference uses samples to summarize $p(f(\theta))$\footnote{In a Bayesian context, the measure of interest is the posterior $p(f(\theta) \mid y)$, but MCMC is more generally applicable to any distribution of $f(\theta)$. To keep our presentation general, we write $p(f(\theta))$, with the understanding that it might be conditional on data.} and typically combines a measure of a central tendency, such as the mean and median, with a measure of uncertainty, such as the variance and 90\% interval.
 
  \item[] {\bf Task 2. Computation} of $\mathbb E (f(\theta))$ or quantiles of 	$f(\theta)$ with respect to the measure $p$.
  
\end{itemize} 
There is overlap between the two tasks: summarizing $p(f(\theta))$ with the mean and variance essentially requires constructing Monte Carlo estimates of $\mathbb E f(\theta)$ and $\mathbb E \left (f(\theta)^2 \right)$.
This is what is commonly done in Bayesian statistics, where our goal is inferential.
How much we learn about $f(\theta)$ is limited by both the Monte Carlo standard error (MCSE) and the posterior standard deviation.
It is useful for the MCSE to be small relative to the posterior standard deviation \textit{to some extent}: at some point further decreasing the MCSE no longer improves our understanding of $f(\theta)$, because the uncertainty is dominated by the posterior standard deviation and moreover limited information we have from our model and data. 
In other problems, the goal is not to learn about $f(\theta)$ but to actually compute $\mathbb E \left (f(\theta) \right)$, as is the case in statistical physics, where expectations correspond to physically meaningful quantities \citep{Landau:2009}.
Then the uncertainty in our results stems from the MCSE only, which we can work harder to reduce. \\

\begin{adjustwidth}{15pt}{15pt}
\noindent
\subsubsection*{\textit{Example from epidemiology: the number of iterations depends on the inferential goal.}}

To illustrate this point, we provide an example from epidemiology.
Our goal is to characterize the dynamics of an influenza A (H1N1) outbreak at a British boarding school.
The data consists of the daily number of students in bed over two weeks and is available in the \texttt{R} package \texttt{outbreaks} \citep{outbreaks}.
We fit a negative binomial distribution parameterized by a susceptible-infected-recovered (SIR) model.
This pedagogical example confronts us with several practical questions and serves as the foundation for many models used to study the Covid-19 outbreak; see \citet{Grinsztajn:2021} for more details.
An epidemiologist can learn many things from such an analysis.
Here we focus on the recovery time $T_\text{R}$, which is derived as a function of the model parameters.
With what precision should we estimate this quantity? 
How much computation should we invest in our Bayesian inference?

We fit the model using the out-of-the-box HMC sampler provided by \texttt{Stan}, which runs 1000 warmup iterations, which we discard, and 1000 sampling iterations.
We run 4 chains in parallel initialized at random draws from the prior.
Table~\ref{tab:recovery} reports the posterior median and 90\% posterior interval (5$^\text{th}$ to 95$^\text{th}$ posterior quantiles) for $T_\text{R}$, obtained across 5 runs of MCMC.
Each run entails 4 Markov chains with 1000 warmup and 1000 sampling iterations, but each time we use a different seed.
All 5 runs produce consistent estimates of the posterior median at least within 2 significant digits.
From one run to the other we get slight disagreements for estimates of the 5$^\text{th}$ and 95$^\text{th}$ posterior quantiles, which suggests that, even after generating thousands of samples, the precision is only with one digit.
This is not surprising: tail quantities, such as extreme quantiles, are more difficult estimate than central quantities such as the median.
This is not a concern for this particular application, since we do not need a precise estimate of the recovery time.
Medical staff only need to know that symptoms will be experienced most likely for 2 days, perhaps 1 or 3 days in some unlikely cases.

\begin{table}
  \begin{center}
  \renewcommand{\arraystretch}{1.5}
  \begin{tabular}{rr}
  Using 4000 draws & Using 40 draws \\\hline
   1.85 [1.61, 2.12] & 1.79 [1.45, 2.18] \\
   1.85 [1.60, 2.11] & 1.82 [1.59, 2.09] \\
   1.85 [1.62, 2.11] & 1.80 [1.62, 2.04] \\
   1.85 [1.62, 2.11] & 1.86 [1.66, 2.25] \\
   1.85 [1.62, 2.12] & 1.84 [1.65, 2.10]
  \end{tabular}
  \end{center}
  \caption{\textit{Estimated posterior median and 90\% interval for the recovery time $T_\text{R}$ from an influenza infection.
  Results from multiple MCMC runs with (left) a long sampling phase and (right) a short sampling phase.}}
  \label{tab:recovery}
\end{table}

If we wanted a better understanding of a patient's recovery time, running more simulations would not be helpful. 
The uncertainty in our inference comes not from the MCSE but from the posterior standard deviation, which is in turn driven by the limited information in our data and the natural variability among patients.
A next step might be to construct a more sophisticated model with patient-level predictors, and collect more data to fit this model.

With regard to the quality of our Monte Carlo estimator, we may wonder if \texttt{Stan}'s defaults were not overkill for this problem.
What if the sampling phase only entailed 10 iterations rather than 1000?
With 4 chains, this would mean generating a mere 40 samples, which in our academic zeitgeist seems unacceptable.
Looking at the Monte Carlo estimates thus obtained (Table~\ref{tab:recovery}) we see that the results exhibit more variance.
Yet this much cruder estimate provides just as much medical insight as the more precise calculation, and all that for much less computation.
Accounting for the chains' autocorrelation, this cheap estimate has an effective sample size (ESS) of $\sim$20, which agrees with MacKay's (in)famous comment that an ESS as small as a dozen can be sufficient \citep[chapter 29]{Mackay:2003}, at least for the purpose of doing inference ({\bf task 1}).
In the next section, we will continue this example and discuss how much computation we need to assess the quality of our Monte Carlo estimators.
\end{adjustwidth}

This example is somewhat provocative (but only somewhat).
Depending on the application, we may require more precise estimates which warrant a long sampling phase.
Drawing from our own experience, we once generated 12000 sampling iterations for a genomic study \citep{Piironen:2017, Margossian:2020}.
One of our goals was to identify explanatory predictors, which required characterizing a high-dimensional and multimodal posterior distribution, and estimating extreme quantiles.\footnote{Admittedly we did not conduct a careful analysis to see how much our scientific conclusions would change if we used a shorter sampling phase.
We did, however, find that the results produced by a fast variational inference approximation were less accurate and did change our insights in a meaningful manner.}
Studies in statistical physics, where {\bf task 2} (computation of expectation values) is the goal rather than a means to {\bf task 1} (inference), can also justify deploying a large amount of computation.
Another example to muse about is Buffon's needle experiment to estimate $\pi$ \citep{Laplace:1812}.\footnote{
  Buffon himself did not try to estimate $\pi$, and it was Laplace who proposed to use a Monte Carlo estimator based on Buffon's study of a dropping needle; see \citet{Badger:1994}.
}
Granted, this is direct Monte Carlo, without any Markov chain, but the point still stands: the number of times we drop the needle may vary greatly depending on the number of digits with which we want to estimate $\pi$.
All in all, understanding the precision needed from our Monte Carlo estimators should lead to more computationally sensible deployments of MCMC, which may depart from general recommendations and software defaults.

When MCMC is used to fit a model, there can be many quantities of interest.
The same reasoning as above applies, with the added complication that we should now have a sense of how the inferences will be used.
For example, when training a Bayesian neural network, our goal is not to do inference on the weights of the network, rather to make accurate and well-calibrated predictions.
Then our intent should be bent on summarizing the posterior distribution of the predictions themselves, or perhaps even some decision based on those predictions, and not on the weights.

The required accuracy of MCMC also depends on the stage of data analysis we find ourselves in.
In Bayesian workflow, model development is an iterative process during which we loop between model building, inference, and model criticism \citep{Box:1987, Blei:2014, Gelman:2020}.
In the early stages of model development, we are trying to identify shortcomings in our model---anything ranging from an implementation error to more substantial flaws in our modeling assumptions.
Many limitations can be identified using a short run of MCMC, rather than waiting hours, days, or weeks to construct long Markov chains.
As we refine our model, we use more precise inference, and additional computational effort can be expanded for the polished models.

\section{Reliability goals}

The accuracy of a Monte Carlo estimator is in general unknown and must be estimated from the simulations themselves.
Returning to our example from epidemiology, 40 sampling iterations provide more than sufficiently accurate estimators but we would not be able to confirm this without either running more chains or running longer chains.
Tacitly, we took comfort in how similar our results were to what we obtained with 4000 samples.

\begin{figure}

  \begin{center}
  \includegraphics[width=4.5in]{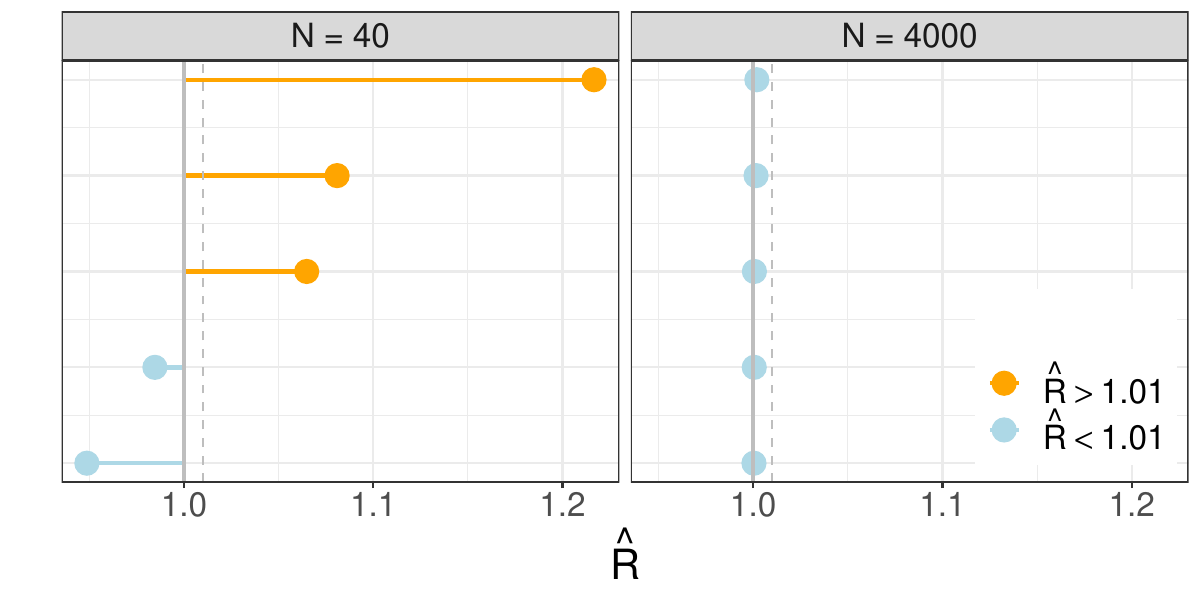}
  \end{center}
  
  \caption{\textit{$\widehat R$ for the recovery time $T_\text{R}$ from an influenza infection.
  We report the results from multiple MCMC runs with (left) a short sampling phase and (right) a long sampling phase.
  With a short sampling phase, $\widehat R$ is noisy and for most seeds incorrectly reports non-convergence, i.e. $\widehat R > 1.01$.}}
  \label{fig:recovery_rhat}
\end{figure}

Determining if our estimators are reliable entails 
\begin{itemize}
  \item[] (i) assessing approximate convergence of the Markov chains, which is an indirect way of checking that the bias of our estimators is small,
  \item[] (ii) estimating the variance of the estimators.
\end{itemize}
For instance, we may use $\widehat R$ \citep{Gelman:1992, Vehtari:2021} for item (i), and an estimator of ESS, which we denote $\widehat{\text{ESS}}$, for item (ii).
For the latter, we use the $\widehat {\text{ESS}}$ estimator from \citet{Vehtari:2021}, which combines autocorrelation functions and estimates of the posterior variance using multiple chains; see also \citet{Gelman:2013, Geyer:1992}.
%
These estimators measure properties of the MCMC sampler, based on one realization (or seed) of the stochastic algorithm,
and so it instructive to compare the estimates between runs (Table~\ref{tab:recovery_diagnostics} and Figure~\ref{fig:recovery_rhat}).
When building long chains, $\widehat R$ is systematically below 1.01, the threshold recommended by \citet{Vehtari:2021} to establish approximate convergence.
On the other hand, with short chains, $\widehat R$ varies between runs and we do not consistently detect convergence. From experimenting with longer warmup phases, we know that for this problem the Markov chains are nearly stationary after a warmup phase of 1000 iterations, even if $\widehat R$ fails to detect it.
When computed using 40 samples, $\widehat R$ is noisy and furthermore biased towards reporting a lack of convergence \citep{Margossian:2023}.
Taking $\widehat R$ at face value and concluding the Markov chains have not converged, we would then question the relevance of the ESS as a measure of the Monte Carlo error, since the ESS is a property of stationary Markov chains and does not account for bias.


\begin{table}
    \begin{center}
    \renewcommand{\arraystretch}{1.5}
    \begin{tabular}{r l r r r r r}
    ${\bf N = 4000}$ & $\widehat R$ & 1.00 & 1.00 & 1.00 & 1.00 & 1.00 \\
      & $\widehat{\text{ESS}}$ & 2659 & 2818 & 2698 & 2457 & 2613 \\
      \rowcolor{lightgray} ${\bf N = 40}$ & $\widehat R$ & 1.22 & 1.08 & 0.95 & 1.06 & 0.98 \\
      \rowcolor{lightgray} & $\widehat{\text{ESS}}$ & 20 & 20 & 20 & 20 & 20
    \end{tabular}
    \end{center}
    \caption{\textit{$\widehat R$ and estimated ESS for the recovery time $T_\text{R}$ from an influenza infection.
    We report the results from multiple MCMC runs with (top) a long sampling phase and (bottom) a short sampling phase.
    When using $N = 40$, the $\widehat R$ and ESS estimators are noisy.}}
    \label{tab:recovery_diagnostics}
\end{table}

The variance and inaccuracy of our error estimators adds a substantial difficulty, which we should see as a third and distinct challenge:
\begin{itemize}

  \item[] {\bf Task 3.} Build {\bf diagnostics} to check the accuracy of our Monte Carlo estimates.

\end{itemize}
This task is particularly tricky, in that it may provide endless justification for skepticism and using more computation.
A common example is the fear of a hidden mode which the chains have failed to find, and so we should keep running the sampler in the off chance this mode exists and that one of the Markov chains stumbles into it.

Setting aside the possibilities of dragons in unexplored regions of probability space, \citet{Vehtari:2021} provide a general recommendation to aim for a \textit{measured} ESS of 100 to ensure good estimates of $\widehat R$ and the ESS, and accordingly \texttt{Stan} warns you that its diagnostics may not be reliable if $\widehat {\text{ESS}} < 100$.
All this does not necessarily contradict the advice by \citet{Mackay:2003} that a sample size of a dozen may suffice, which applies to the true ESS, assumed to be measured perfectly, rather than the measured ESS.
Here it is essential to clearly distinguish the three tasks and recognize that, in general, there is a gap between the number of iterations required to do well in {\bf task 2} (computation) and in {\bf task 3} (diagnostics), with the latter being more expensive.
Reducing this gap constitutes an important step towards decreasing the run time and increasing the reliability of MCMC.


\section*{Part II: Theoretical motivation, warnings, and open problems}

We now examine the question of MCMC computation from a more theoretical perspective.
We provide justification for our general recommendations (Section~\ref{sec:recommendations}), as well as reasons to sometimes depart from these in light of recent methodological and hardware advances.

\section{Error in Monte Carlo estimators}

Consider a state space $\Theta$ over which the target distribution $p$ is defined.
Moving on, we focus on estimating expectations $\mathbb E \left( f(\theta) \right)$, where $\theta \in \Theta$ and $f$ maps $\theta$ to a univariate scalar.
Usually, we are interested in multiple such scalar summaries.
Let $\bar f$ be our Monte Carlo estimator, obtained by averaging the MCMC samples ${\bf f} = \left ( f(\theta^{(1)}), f(\theta^{(2)}), \cdots, f(\theta^{(N)}) \right)$,
\begin{equation}
  \bar f = \frac{1}{N} \sum_{n = 1}^N f \left (\theta^{(n)} \right),
\end{equation}
and let $\Gamma_N$ be the distribution from which this estimator is sampled,
\begin{equation}
  \bar f \sim \Gamma_N.
\end{equation}
%
%
When determining the control parameters of MCMC, we aim to control the expected squared error, which decomposes into a squared bias and a variance component
\begin{equation}
  \mathbb E_{\Gamma_N} \left (\bar f - \mathbb E_p f(\theta) \right)^2 
    = \underbrace{\left (\mathbb E_{\Gamma_N} \bar f - \mathbb E_p f(\theta) \right)^2}_{\text{Squared Bias}} + \underbrace{\text{Var}_{\Gamma_N} \bar f}_{\text{Variance}},
\end{equation} 
where the subscripts denote, for clarity, the distribution with respect to which the expectation value is taken.
The bias is due to the fact we start the Markov chain at $p_0$ rather than $p$, and the variance arises because we compute sample estimates with a finite number of samples.

\begin{adjustwidth}{15pt}{15pt}
\noindent
\subsubsection*{\textit{Example: Langevin diffusion approximation of MCMC.}}
Consider a Langevin diffusion that evolves from the univariate distributions $p_0 = \text{normal}(\mu_0, \sigma_0)$ to $p = \text{normal}(\mu, \sigma)$, and suppose $f$ is the identity function, $f(\theta) \equiv \theta$.
The Langevin diffusion provides a continuous approximation of MCMC and the discrete number of iterations is now replaced by the continuous time $T$ \citep{Gelman:1997, Roberts:1998}.
The Monte Carlo estimator is given by the integral,
\begin{equation}
  \bar \theta = \frac{1}{T} \int_0^T \theta^{(t)} \text d t.
\end{equation}
What makes this example particularly enlightening is that the distribution of the sample $\theta^{(t)}$ can be written analytically,
\begin{equation} \label{eq:langevin}
  \theta^{(t)} \sim \text{normal} \left (\mu + (\mu_0 - \mu) e^{-t}, \sqrt{\sigma^2 + (\sigma_0^2 - \sigma^2) e^{-2t}} \right).
\end{equation}
%
In this example the bias decays at an exponential rate, although for $t < \infty$ it never hits 0.
Similarly standard MCMC algorithms cannot produce unbiased estimators, nor will finite computation return a single sample from the target distribution $p$.
This is fine: all we need is a small enough bias, just as achieving approximate stationarity suffices.

\noindent
{\bf Remark.} \textit{On a computer with finite precision, we may not be able to distinguish the bias from 0 when $T$ is large.
In practice, we often do not need the bias to be 0 within floating point precision, rather it should be sufficiently small next to the Monte Carlo standard error and posterior standard deviation.
}
\end{adjustwidth}

Typically, we discard the first iterations as part of a warmup phase (also termed ``burn-in'').
The primary task of the warmup is to allow the bias of the Markov chain to decay before we start sampling.
The primary goal of the sampling phase is to then reduce the variance, which would still exist even in the ideal circumstance where the Markov chains are stationary.

This dichotomy suggests a bias-variance tradeoff, when choosing the number of samples to discard in warmup, a problem which has been tackled in the context of Stein methods; see \citet{South:2021} and references therein.
But the story admits some nuances.
Typically the variance also decays during warmup, especially if $\sigma_0 > \sigma$ and also because of the Markov chain's drift (averaging samples with different means increases the variance of $\bar f$).
This additional variability, which we later define as the \textit{nonstationary variance}, plays a key role in diagnosing whether the Markov chains are close to their stationary distribution.
Similarly, the bias continues to decay during the sampling phase.
However, once the drift becomes minor and the initial variance less influential, there is a net benefit, in terms of expected squared error, to including still biased samples rather than discarding them.

While we ultimately care about the expected squared error, we tend to tolerate variance more than bias because of {\bf task 3} (diagnostics).
That is, we can estimate the variance of our estimator, usually reported via the Monte Carlo standard error (MCSE) or the ESS, while no good estimator of the bias exists.
The hope is then that the bias is negligible next to the measured variance and the squared error well characterized by the estimated variance.
This will be the case for nearly stationary Markov chains, and so a common strategy is to first assess approximate convergence as a proxy for bias decay.

In some cases, it is possible to construct unbiased estimators.
In our view, the main limitation of MCMC is not its bias but rather that we cannot guarantee that this bias has vanished.
Therefore techniques which entirely remove the bias are primarily useful for {\bf task 3} (diagnostics).
The \textit{unbiased MCMC framework} \citep{Jacob:2020} is in that respect promising, although the construction of the unbiased estimator can, for certain problems, require a long warmup phase (or \textit{coupling time}).
This ties back to the question of how much additional computation can we afford to pay, on top of what might be required to achieve an acceptable precision, to assess the reliability of our estimators.

\section{Variance of Monte Carlo estimator}

Beyond the standard advice to run the chain longer, there exist several techniques to reduce the variance of our Monte Carlo estimators.
These techniques can help us achieve our goals with fewer sampling iterations. 
We review recent work on the use of massive parallelization to reduce variance \citep{Lao:2020}.
Another variance reduction technique of interest is control variates.
We do not discuss this method here, and direct the readers to some references \citep[e.g][]{South:2023, Mira:2013}.

\subsection*{Effective sample size}

The length of the chain is a poor proxy for how precise an estimator is, given the chain's autocorrelation drastically impacts how much the variance decreases with each iteration.
A useful measure is the ESS, which can be estimated using autocorrelation functions.
We here provide two ways to interpret the ESS for stationary Markov chains (and, for our purpose, approximately stationary Markov chains):
\begin{itemize}
  \item It is the number of independent samples from $p(\theta)$ we would require to construct an estimator with the same standard deviation as our Monte Carlo estimator $\bar f$.
  
  \item It is a ratio of variances, $\text{ESS} = \text{Var}_p f(\theta) / \text{Var}_\Gamma \bar f$.
  For independent samples, this ratio becomes the number of samples $N$.
\end{itemize}
Understanding the ESS required for an application can help us determine if the sampling phase is sufficiently long.
The ESS can be tied to the number of significant digits in our Monte Carlo estimator, after scaling by the standard deviation.
Precise recommendations require determining with what probability we would allow a significant digit to vary between MCMC runs and we refer the reader to \citet{Vats:2019} and \citet{Vehtari:2022}.
As a rule of thumb, achieving an MCSE which is 0.1 the posterior standard deviation requires an ESS of 100; adding another significant digit requires an ESS of 10,000.
The ESS provides a scale free measure of our Monte Carlo estimator's precision and complements the non-scale free MCSE.
Crucially the ESS can vary between quantities of interest, even when these are evaluated using the same MCMC samples: for example, anticorrelated samples produce precise estimates of the first moment but poor estimates of the second moment.
Therefore the ESS should be examined for all quantities of interest.

\subsection*{Choosing a sampler and tuning it to reduce variance}

Given a fixed number $N$ of sampling iterations, the  ESS can fluctuate dramatically depending on the MCMC algorithm and the target distribution $p$.
It is critical to chose an appropriate sampler to handle the pathologies that can arise in a target distribution, including high dimension, correlation, uneven curvature, multimodality, and more.
How to chose a sampler is a broad topic.
We usually start with HMC which is general purpose and benefits from several high performance implementations, for example in \texttt{Stan}, \texttt{PyMC}, and more.
HMC is not universal: it cannot directly handle discrete parameters and will fare poorly with well separated modes and multiscale distributions, and so other samplers can be invoked to tackle a particular problem.

Even within a class of MCMC algorithms, the chain's autocorrelation can be sensitive to certain tuning parameters, which we must either set manually or learn during the warmup phase.
Self-tuning algorithms are a critical component of Bayesian workflow, since they allow us to move from one model to the other without having to revise our inference algorithm, and are also much more accessible to the broader scientific community.
In \texttt{Stan}, we employ dynamic HMC \citep{Betancourt:2018}, based on the no-U-turn sampler (NUTS) \citep{Hoffman:2014}, which learns HMC's tuning parameters to maximize the expected squared jump distance.
After warmup, we freeze the tuning parameters of HMC because continuously adapting the tuning parameters can produce the wrong stationary distribution \citep{Andrieu:2008}.
Certain adaptation strategies are not susceptible to this phenomenon and continuously target the right stationary distribution while adapting \citep[e.g.,][]{Gilks:1994, Hoffman:2022}.

\subsection*{The many-short-chains regime}  \label{sec:many-short-chains}

\begin{figure}
\begin{center}
\resizebox{4.5in}{2.25in}{%
\begin{tikzpicture}
    [
       Orange/.style={circle, draw=orange!, fill=green!0, thick, minimum size=10},
       Blue/.style={circle, draw=charlesBlue!, fill=green!0, thick, minimum size=10},
       Gray/.style={circle, draw=gray!, fill=green!0, dashed, minimum size=10},
       Empty/.style={, draw=white!, fill=green!0, minimum size=0mm}
    ]

    \node[Empty] (p0) at (0, 0.5) {$p_0$};
    \node[Empty] (phat) at (5, 0.5) {$\hat p \approx p$};

    \node[Empty] (time) at(2, -3) {$\underbrace{\hspace{4.5cm}}$};
    \node[Empty] (warmup) at (2, -3.5) {\small warmup phase};

     \node[Empty] (time) at(7, -3) {$\underbrace{\hspace{4.5cm}}$};
    \node[Empty] (warmup) at (7, -3.5) {\small sampling phase};
    
    \node[Orange] (theta0) at (0, 0) {};
    \node[Orange] (theta1) at (1, 0) {};
    \node[Orange] (theta2) at (2, 0) {};
    \node[Orange] (theta3) at (3, 0) {};
    \node[Orange] (theta4) at (4, 0) {};
    \node[Blue] (theta5) at (5, 0) {};
    \node[Gray] (theta6) at (6, 0) {};
    \node[Gray] (theta7) at (7, 0) {};
    \node[Gray] (theta8) at (8, 0) {};
    \node[Gray] (theta8) at (9, 0) {};

    \node[Orange] (1theta0) at (0, -0.5) {};
    \node[Orange] (1theta1) at (1, -0.5) {};
    \node[Orange] (1theta2) at (2, -0.5) {};
    \node[Orange] (1theta3) at (3, -0.5) {};
    \node[Orange] (1theta4) at (4, -0.5) {};
    \node[Blue] (1theta5) at (5, -0.5) {};
    \node[Gray] (1theta6) at (6, -0.5) {};
    \node[Gray] (1theta7) at (7, -0.5) {};
    \node[Gray] (1theta8) at (8, -0.5) {};
    \node[Gray] (1theta8) at (9, -0.5) {};

    \node[Orange] (2theta0) at (0, -1) {};
    \node[Orange] (2theta1) at (1, -1) {};
    \node[Orange] (2theta2) at (2, -1) {};
    \node[Orange] (2theta3) at (3, -1) {};
    \node[Orange] (2theta4) at (4, -1) {};
    \node[Blue] (2theta5) at (5, -1) {};
    \node[Blue] (2theta5) at (5, -1) {};
    \node[Gray] (2theta6) at (6, -1) {};
    \node[Gray] (2theta7) at (7, -1) {};
    \node[Gray] (2theta8) at (8, -1) {};
    \node[Gray] (2theta8) at (9, -1) {};
    
    \node[Empty] (manyChains) at (0, -1.65) {$\vdots$};
 
    \node[Orange] (Mtheta0) at (0, -2.5) {};
    \node[Orange] (Mtheta1) at (1, -2.5) {};
    \node[Orange] (Mtheta2) at (2, -2.5) {};
    \node[Orange] (Mtheta3) at (3, -2.5) {};
    \node[Orange] (Mtheta4) at (4, -2.5) {};
    \node[Blue] (Mtheta5) at (5, -2.5) {};
    \node[Blue] (Mtheta5) at (5, -2.5) {};
    \node[Gray] (Mtheta6) at (6, -2.5) {};
    \node[Gray] (Mtheta7) at (7, -2.5) {};
    \node[Gray] (Mtheta8) at (8, -2.5) {};
    \node[Gray] (Mtheta8) at (9, -2.5) {};
   
    \draw [-latex](theta0) -- (theta1);
    \draw [-latex](theta1) -- (theta2);
    \draw [-latex](theta2) -- (theta3);
    \draw [-latex](theta3) -- (theta4);
    \draw [-latex](theta4) -- (theta5);

    \draw [-latex](1theta0) -- (1theta1);
    \draw [-latex](1theta1) -- (1theta2);
    \draw [-latex](1theta2) -- (1theta3);
    \draw [-latex](1theta3) -- (1theta4);
    \draw [-latex](1theta4) -- (1theta5);
    
    \draw [-latex](2theta0) -- (2theta1);
    \draw [-latex](2theta1) -- (2theta2);
    \draw [-latex](2theta2) -- (2theta3);
    \draw [-latex](2theta3) -- (2theta4);
    \draw [-latex](2theta4) -- (2theta5);

    \draw [-latex](Mtheta0) -- (Mtheta1);
    \draw [-latex](Mtheta1) -- (Mtheta2);
    \draw [-latex](Mtheta2) -- (Mtheta3);
    \draw [-latex](Mtheta3) -- (Mtheta4);
    \draw [-latex](Mtheta4) -- (Mtheta5);

    \end{tikzpicture}
    }
\end{center}
\caption{\textit{The many-short-chains regime of MCMC. Increasing the number of Markov chains reduces the variance of our Monte Carlo estimators, meaning we can achieve a desired precision with a short sampling phase. However the bias cannot be averaged out and so each chain must still be properly warmed up.}}
\label{fig:many-short-chains}
\end{figure}
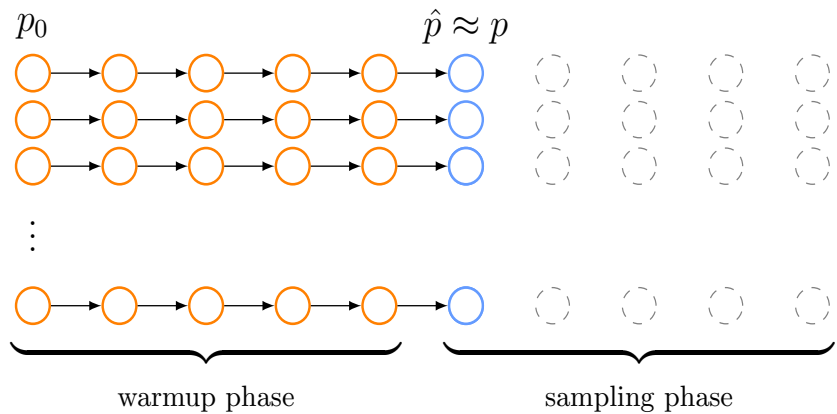

A natural way to increase the number of samples is to run more chains.
It has long been recognized that we can trade the length of the sampling phase for the number of chains when controlling the variance of $\bar f$, and furthermore that the additional computation time can be mitigated by running all chains in parallel \citep{Rosenthal:2000}.
There are some difficulties we need be mindful of and, in general, increasing the number of chains still increases the run time, even with parallelization.
Some of these difficulties include potentially failing machines \citep{Rosenthal:2000}, algorithms with a heavy control flow \citep{Lao:2020, Hoffman:2021} and Markov chains with wildly varying run times \citep{DuChe:2023}.


Over the past decade, the balance of computing power has shifted from CPUs to GPUs.
A GPU boasts thousands of cores,\footnote{
  The GPU core has different qualities than its CPU counterpart, which can be problematic for certain operations.
  For example, sequential calculations, such as numerically solving a differential equation, work much better on a CPU core than a GPU one.} which raises the possibility of running this many chains in parallel.
To do this efficiently, hardware alone is not enough, and several challenges must be overcome on the algorithmic front.
Here we refer the reader to the recent literature on GPU-friendly MCMC \citep[e.g.,][]{Lao:2020, Hoffman:2021}.
Programs such as \texttt{TensorFlow Probability} in \texttt{Python} are expressly designed to run hundreds or even thousands of Markov chains on a GPU \citep{tfp:2023}.

With each additional Markov chain, we further reduce the variance of $\bar f$.
Equivalently, we reduce the number of sampling iterations required to achieve a target variance (Figure~\ref{fig:many-short-chains}).
In the next section, we will discuss how running multiple chains can also affect the warmup phase but for now, we assume the warmup of each chain is sufficiently long and that the Markov chains are (approximately) stationary.
Given enough Markov chains, the sampling phase can be arbitrarily short, potentially containing a single iteration per chain.
For example, if our target ESS is, say, 1000, then warming up 1000 Markov chains and generating one sampling iteration per chain produces the target ESS.
Furthermore, this target ESS is achieved \textit{for every variable} $f(\theta)$, and if the chains are independent, we also have a central limit theorem (CLT) along the number of chains, rather than along the number of sampling iterations.
However this CLT, unlike the MCMC CLT, does not ensure that the bias vanishes---that is why we we still need a warmup phase---rather it allows us to characterize the Monte Carlo error due to variance. 

If we keep running the chains longer, the ESS further decreases for a relatively small computational cost but this might be overkill if we have already achieved the wanted precision.
Similarly, running more chains would not help us attain our target precision faster.
In this sense, the requisite ESS sets an upper bound on how many Markov chains it is useful to run for the purpose of reducing variance.
If we only require an ESS of a few dozen, then there is no use exploiting all 10,000 cores available on a GPU to run that many chains.
These cores can always be put to work elsewhere, perhaps to do within-chain parallelization \citep[e.g.,][]{Lee:2010, Cesnovar:2020}.
One appeal of running many short chains is that this can sidestep the problem of choosing the length of the sampling phase,
since the variance reduction for every variable can be entirely handled by the number of chains.
The warmup phase then dominates the computation, and algorithmic choices can solely focus on building a sampler with a fast decaying bias.

We will see in the next section that the use of cross-chain adaptation can also lead to faster bias decay during the warmup phase as we increase the number of chains. In addition to bias and variance considerations, which relate to {\bf task 1} (inference) and {\bf task 2} (computation), running many chains can also improve {\bf task 3} (diagnostics), for instance by increasing the probability of finding multiple modes and reducing the variance of statistics such as $\widehat R$.

\section{Bias of MCMC and approximate convergence}

Bias reduction has played a less prominent role than variance reduction in the MCMC literature.
For many problems, we can expect the bias to decay faster than the variance as the length of the Markov chain increases.
In the earlier example of the Langevin diffusion (Eq~\ref{eq:langevin}), the bias decreases at an exponential rate.
On the other hand, the variance decreases linearly with the time $T$ of the process.
These types of examples may explain why the empirical performance of MCMC algorithms is mostly reported in terms of ESS per operation, which, above all, tells us how well an algorithm performs once approximate stationarity is achieved.

More recent papers, notably on the topic of GPU-friendly samplers, experimentally examine the rate at which the bias decreases \citep[e.g.,][]{Hoffman:2022}, perhaps foreshadowing a shift in paradigm. 
Indeed the recent development of general-purpose variance reduction techniques, such as control variates or plainly running many chains, means our target variance can be attained with a much shorter sampling phase than what would traditionally be prescribed.
In general these techniques do not reduce the bias of our estimators.
Hence, for problems amiable to variance reduction techniques, the bias is the primary computational bottleneck and the main requirement for an MCMC algorithm is to converge quickly to its stationary distribution.

\subsection*{Choosing a sampler and tuning it to reduce bias}

Many tuning criterions, such as maximizing the expected squared jump distance \citep{Pasarica:2010}, tacitly focus on the \textit{asymptotic variance} of MCMC, once the Markov chains are (approximately) stationary.
A complementary measure of efficiency is the \textit{speed of convergence}, usually expressed in terms of the \textit{total variational distance} ($D_\text{TV}$) between the approximate distribution $\hat p^{(t)}$ of $\theta^{(t)}$ and the stationary distribution $p$,
\begin{equation}
  D_\text{TV} = \sup_{A \subseteq \Theta} \left | \int_{\theta \in A} \hat p^{(t)}(\theta) - p(\theta) \text d \theta \right |.
\end{equation}
We then define the \textit{relaxation time} as the number of iterations required to move from an initial distribution $p_0$ to $\hat p^{(t)}$ such that $D_\text{TV} \left ( \hat p^{(t)}, p \right) \le \epsilon$ for some pre-defined $\epsilon > 0$.
The total variation distance is not a perfect proxy for the bias, but it provides an upper bound for the absolute bias when estimating the expectation value of a bounded function $f$ \citep[e.g][Proposition 3]{Roberts:2004}.
Specifically, given two scalars $a < b$,
\begin{equation}
    D_\text{TV} = \frac{1}{b - a} \sup_{f: \Theta \to [a, b]} \left | \mathbb E_{\hat p^{(t)}} \left ( f(\theta) \right) - \mathbb E_p \left ( f(\theta) \right)  \right |.
\end{equation}
%
We might argue that all functions $f$ are bounded, given finite machine precision.
In this sense, the speed of convergence and relaxation time provides a conservative indication of how efficient a sampler is during the warmup phase, i.e. the rate per iterations at which we reduce the bias for all estimators of interest.
The asymptotic variance, on the other hand, tells us how efficient the MCMC sampler is during the sampling phase. 

Asymptotic efficiency and speed of convergence may not prescribe the same optimal tuning of an MCMC algorithm \citep{Besag:1993, Mira:2001}.
We may in theory favor one efficiency goal over the other, depending on whether we expect bias or variance reduction to be the main computational bottleneck.

\begin{adjustwidth}{15pt}{15pt}
\noindent
\subsubsection*{\textit{Example: the transition kernel that minimizes the relaxation time does not minimize the asymptotic variance.}} 

Consider a finite discrete space.
The choice that minimizes $D_\text{TV}$ is trivially the transition kernel that generates an independent sample from $p$.
With such an ``oracle'' choice, the relaxation time is a single iteration.
Now suppose we are at a point $\theta^{(t)}$.
The oracle generates with a nonzero probability $\theta^{(t + 1)} = \theta^{(t)}$, but when estimating $\mathbb E (\theta)$, it is often preferable to pick a transition kernel that always moves to another point, such that $\theta^{(t + 1)} \neq \theta^{(t)}$ and the samples are anti-correlated.
Therefore, the choice that optimizes speed of convergence differs from the one that would minimize the asymptotic variance.
\end{adjustwidth}

In practice, we find that algorithms which are asymptotically efficient also perform well in terms of speed of convergence.
We are also not aware of practical tuning criterions, which do not involve oracle knowledge and still explicitly target speed of convergence.
Developing such methods---and whether it is useful to do so---constitutes an open research question.

\begin{figure}
  \begin{center}
    \includegraphics[width=3in]{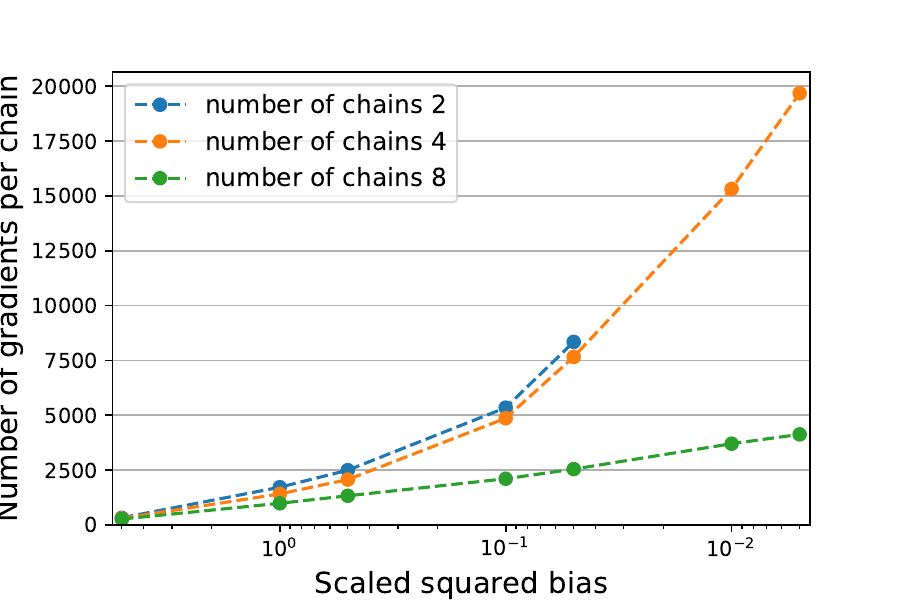}
    \caption{\textit{Number of gradient evaluations per chain required for ChEES-HMC to achieve a target bias across all $d = 501$ dimensions of an ill-conditioned Gaussian.
    Gradient evaluation is the dominant computational operation for ChEES-HMC.
    Due to cross-chain adaptation, increasing the number of chains can lead to a faster bias decay.
    When using 2 chains, we do not achieve a bias below $10^{-2}$ after 1000 warmup iterations.}}
    \label{fig:ChEES-HMC}
  \end{center}
\end{figure}

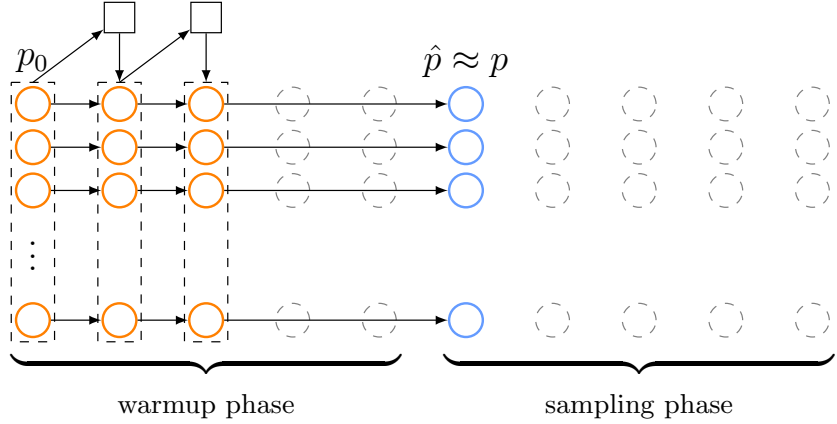
\begin{figure}
  \begin{center}
  \resizebox{4.5in}{2.25in}{%
  \begin{tikzpicture}
    [
       Orange/.style={circle, draw=orange!, fill=green!0, thick, minimum size=10},
       Blue/.style={circle, draw=charlesBlue!, fill=green!0, thick, minimum size=10},
       Gray/.style={circle, draw=gray!, fill=green!0, dashed, minimum size=10},
       Square/.style={rectangle, draw=black!, minimum size=10},
       Empty/.style={, draw=white!, fill=green!0, minimum size=0mm}
    ]

    \node[Square] (c1) at (1, 1) {};
    \node[Square] (c2) at (2, 1) {};

    \node[Empty] (p0) at (0, 0.5) {$p_0$};
    \node[Empty] (phat) at (5, 0.5) {$\hat p \approx p$};

    \node[Empty] (time) at(2, -3) {$\underbrace{\hspace{4.5cm}}$};
    \node[Empty] (warmup) at (2, -3.5) {\small warmup phase};

     \node[Empty] (time) at(7, -3) {$\underbrace{\hspace{4.5cm}}$};
    \node[Empty] (warmup) at (7, -3.5) {\small sampling phase};
    
    \node[Orange] (theta0) at (0, 0) {};
    \node[Orange] (theta1) at (1, 0) {};
    \node[Orange] (theta2) at (2, 0) {};
    \node[Gray] (theta3) at (3, 0) {};
    \node[Gray] (theta4) at (4, 0) {};
    \node[Blue] (theta5) at (5, 0) {};
    \node[Gray] (theta6) at (6, 0) {};
    \node[Gray] (theta7) at (7, 0) {};
    \node[Gray] (theta8) at (8, 0) {};
    \node[Gray] (theta8) at (9, 0) {};

    \node[Orange] (1theta0) at (0, -0.5) {};
    \node[Orange] (1theta1) at (1, -0.5) {};
    \node[Orange] (1theta2) at (2, -0.5) {};
    \node[Gray] (1theta3) at (3, -0.5) {};
    \node[Gray] (1theta4) at (4, -0.5) {};
    \node[Blue] (1theta5) at (5, -0.5) {};
    \node[Gray] (1theta6) at (6, -0.5) {};
    \node[Gray] (1theta7) at (7, -0.5) {};
    \node[Gray] (1theta8) at (8, -0.5) {};
    \node[Gray] (1theta8) at (9, -0.5) {};

    \node[Orange] (2theta0) at (0, -1) {};
    \node[Orange] (2theta1) at (1, -1) {};
    \node[Orange] (2theta2) at (2, -1) {};
    \node[Gray] (2theta3) at (3, -1) {};
    \node[Gray] (2theta4) at (4, -1) {};
    \node[Blue] (2theta5) at (5, -1) {};
    \node[Gray] (2theta6) at (6, -1) {};
    \node[Gray] (2theta7) at (7, -1) {};
    \node[Gray] (2theta8) at (8, -1) {};
    \node[Gray] (2theta8) at (9, -1) {};
    
    \node[Empty] (manyChains) at (0, -1.65) {$\vdots$};
 
    \node[Orange] (Mtheta0) at (0, -2.5) {};
    \node[Orange] (Mtheta1) at (1, -2.5) {};
    \node[Orange] (Mtheta2) at (2, -2.5) {};
    \node[Gray] (Mtheta3) at (3, -2.5) {};
    \node[Gray] (Mtheta4) at (4, -2.5) {};
    \node[Blue] (Mtheta5) at (5, -2.5) {};
    \node[Gray] (Mtheta6) at (6, -2.5) {};
    \node[Gray] (Mtheta7) at (7, -2.5) {};
    \node[Gray] (Mtheta8) at (8, -2.5) {};
    \node[Gray] (Mtheta8) at (9, -2.5) {};
   
    \draw [-latex](theta0) -- (theta1);
    \draw [-latex](theta1) -- (theta2);
    \draw [-latex](theta2) -- (theta5);

    \draw [-latex](1theta0) -- (1theta1);
    \draw [-latex](1theta1) -- (1theta2);
    \draw [-latex](1theta2) -- (1theta5);
    
    \draw [-latex](2theta0) -- (2theta1);
    \draw [-latex](2theta1) -- (2theta2);
    \draw [-latex](2theta2) -- (2theta5);

    \draw [-latex](Mtheta0) -- (Mtheta1);
    \draw [-latex](Mtheta1) -- (Mtheta2);
    \draw [-latex](Mtheta2) -- (Mtheta5);

    \draw [-latex, black] (0, 0.25) -- (c1);
    \draw [-latex, black] (c1) -- (1, 0.25);
    \draw [-latex, black] (1, 0.25) -- (c2);
    \draw [-latex, black] (c2) -- (2, 0.25);

    \draw [dashed, black] (-0.25, 0.25) -- (.25, 0.25) -- (.25, -2.75)
      -- (-.25, -2.75) -- (-.25, 0.25);

    \draw [dashed, black] (0.75, 0.25) -- (1.25, 0.25) -- (1.25, -2.75)
      -- (0.75, -2.75) -- (0.75, 0.25);

    \draw [dashed, black] (1.75, 0.25) -- (2.25, 0.25) -- (2.25, -2.75)
      -- (1.75, -2.75) -- (1.75, 0.25);

    \end{tikzpicture}
    }
  \end{center}
  \caption{\textit{Many-short-chains regime using cross-chain adaptation. Using many chains reduces the length of the sampling phase required to achieve a target precision. 
  Cross-chain adaptation pools information between the chains to tune the sampler during the warmup phase.
  For some problems, the improved adaptation means we achieve an acceptable bias with fewer warmup iterations per chain.}}
  \label{fig:many-cross-chains}
\end{figure}

Adaptive algorithms which learn a sampler's tuning parameters in fewer iterations may also converge faster, even if the tuning criterion is based on optimizing the asymptotic variance.
We illustrate this with ChEES-HMC \citep{Hoffman:2021}, a tuning free HMC which targets the expected squared jump distance and uses \textit{cross-chain adaptation}.
In cross-chain adaptation, samples across multiple chains are pooled to tune the sampler's parameters, and so increasing the number of chains can lead to faster adaptation and in turn faster bias decay.
We demonstrate this phenomenon when applying ChEES-HMC to an ill-conditioned Gaussian with dimension $d = 501$.
In ChEES-HMC, the computation cost is dominated by gradient evaluations of $\log p(\theta)$.
We run 1000 warmup iterations using 2, 4, and 8 communicating chains, and report the number of gradient evaluations of $\log p(\theta)$ \textit{per chain} required to achieve a target bias (Figure~\ref{fig:ChEES-HMC}).
As we increase the number of chains, we attain an acceptable bias with fewer gradient evaluations per chain.
With parallelization, this result translates into an optimal warmup phase with a shorter computation time. 
A handful of algorithms now use cross-chain adaptation \citep{Sountsov:2021, Hoffman:2022, Gabrie:2022, Riou-Durand:2023} and for these, running more chains can justify a shorter warmup phase (Figure~\ref{fig:many-cross-chains}).

\subsection*{Assessing approximate convergence with multiple chains}

A general framework to assess convergence is to compare multiple chains \citep{Gelman:1992}.
Each chain is distinguished by its initialization and its seed but neither should have a strong influence on our estimate of $\mathbb E f(\theta)$.
It is therefore natural to check that, subject to varying initializations and seeds, our algorithm still returns the same result.
This idea underlies many standard diagnostics, including visual checks such as comparing trace and density plots generated by different chains, as well as numerical diagnostics such as $\widehat R$ and its variants \citep{Gelman:1992, Brooks:1998, Vehtari:2021, Moins:2022, Vats:2021, Lambert:2022, Margossian:2023}.

A quantity of particular interest is the sample variance of the \textit{per chain Monte Carlo estimator}, obtained by averaging the samples within a single chain.
Let $\theta^{(nm)}$ denote the $n^\text{th}$ sampling iteration in the $m^\text{th}$ chain.
Then the per chain Monte Carlo estimator is
\begin{equation}
  \bar f^{(m)} = \frac{1}{N} \sum_{n = 1}^N f \left(\theta^{(nm)} \right),
\end{equation}
and the final Monte Carlo estimator is obtained by averaging the means of all chains,
\begin{equation}
  \bar f = \frac{1}{M} \sum_{m = 1}^M \bar f^{(m)}.
\end{equation}
We then study the empirical variance,
\begin{equation}
  \widehat B = \frac{1}{M - 1} \sum_{m = 1}^M \left ( \bar f^{(m)} - \bar f \right)^2,
\end{equation}
and check that it is smaller than a predetermined threshold.
Doing so is analogous to the graphical comparisons between chains with trace and density plots (although more information can be inferred from these plots),
and it is exactly what happens when we use the $\widehat R$ convergence diagnostic.

To see this, consider the within-chain variance, defined as the average per-chain sample variance,
\begin{equation}
  \widehat W = \frac{1}{M} \sum_{m = 1}^M \frac{1}{N - 1} \sum_{n = 1}^N \left ( f \left(\theta^{(nm)} \right) - \bar f^{(m)} \right)^2.
\end{equation}
Then we define $\widehat R$ as
\begin{equation}
  \widehat R = \sqrt{\frac{N - 1}{N} + \frac{\widehat B}{\widehat W}}\ ,
\end{equation}
and so checking that $\widehat R \le 1 + \epsilon$ is equivalent to setting a scaled tolerance on $\widehat B$,
\begin{equation}
  \widehat B \lessapprox 2 \epsilon \widehat W + \mathcal O(\epsilon^2).
\end{equation}
This is not the original motivation of $\widehat R$, and there are other useful perspectives on the subject \citep[see][and discussion]{Gelman:1992, Vehtari:2021}.
Here we follow the argument of \citet{Margossian:2023}.  

Convergence diagnostics based on multiple chains largely rely on the variance of $\bar f^{(m)}$, which raises a paradox: how can a variance-based diagnostic tell us whether the warmup phase, whose primary role is to reduce bias, is sufficiently long?
To answer this question, we decompose the MCMC process $\Gamma$ into an initial draw $\theta^{(0)} \sim p_0$ and then a stochastic process $\gamma$, which is the subsequent application of the transition kernels (warmup and sampling included).
Then applying the law of total variance,
\begin{equation}
  \text{Var}_{\Gamma_N} \ \bar f^{(m)} = \underbrace{\text{Var}_{p_0} \left [ \mathbb E_{\gamma_N} (\bar f^{(m)} \mid \theta_0) \right]}_{\text{nonstationary variance}}
    + \underbrace{\mathbb E_{p_0} \left [ \text{Var}_{\gamma_N} (\bar f^{(m)} \mid \theta_0) \right]}_{\text{persistent variance}}.
\end{equation}
The \textit{nonstationary variance} tells us how much the expectation value of our estimator varies with the initialization and provides an intuitive measure of how well the Markov chains forget their starting points.
This term is related to other notions of convergence: as the total variation distance $D_\text{TV}$ and the bias go to 0, so does the nonstationary variance.
\citet{Margossian:2023} show that in the example of a Langevin diffusion evolving from one Gaussian to the other, the squared bias and the nonstationary variance both decay at the same rate $\sim e^{-2t}$,
and so one can be used as a ``proxy clock'' of the other.

The \textit{persistent variance}, named so because it does not go to zero after the Markov chains converge, is of interest when determining whether the sampling phase is sufficiently long.
For stationary Markov chains, the ratio $\widehat W / \widehat B$ provides an estimator of the ESS \citep{Gelman:2003, Vats:2021}, albeit a less stable one than estimators based on autocorrelation functions.

In some sense, $\widehat R$ serves as a diagnostic for both the lengths of the warmup and the sampling phases, and this can be seen as a desirable feature:
it is easier to reason about a single summary quantity than multiple ones.
But if $\widehat R$ is large, it is unclear how to disentangle whether our warmup or our sampling phase is too short.
Another potential limitation is that $\widehat R$ is always large when we use a short sampling phase, even when the error of our final estimator $\bar f$ is small, for example when we run many short chains.
This and other considerations led to the development of a \textit{nested} $\widehat R$, which compares groups of chains initialized at the same starting point and, in doing so, can provide a direct measure of the nonstationary variance, rather than of the total variance \citep{Margossian:2023}.
The role of the convergence diagnostic is then clearly confined to checking the length of the warmup phase; once we establish that the warmup is long enough, we then look at measures of the ESS and the MCSE, and decide whether the sampling phase has a satisfactory length. 

A limitation with diagnostics based on multiple chains is that they essentially rely on problems in our MCMC manifesting as disagreements between chains.
A common concern is that all the chains may be in good agreement but still all be wrong.
For example, they may all gravitate towards the same mode and ignore a hidden mode.
Here we must reason about our diagnostics as estimators (of some relevant quantity), vulnerable to bias and variance, with steps that can be taken to reduce the error:
for instance, running many chains for more iterations increases our ability to detect the hidden mode.

\section{Initial distribution}

Another decision when running MCMC is the choice of initial values.
Here we must balance a few considerations that will in turn accommodate {\bf task 2} (computation) or {\bf task 3} (diagnostics).

With respect to {\bf task 2} (computation), starting with a distribution which is close to the stationary distribution $p$ means we can achieve a target bias with less iterations.
In the Langevin diffusion example, it is clearly desirable to have $(\mu_0, \sigma_0)$ as close as possible to $(\mu, \sigma)$.
We may resort to fast approximations of the target distribution $p$, for example using a Laplace approximation or variational inference.
Recently, \citet{Zhang:2022} showed that \textit{Pathfinder} variational inference can initially produce better approximations faster than HMC-NUTS across a range of Bayesian inference problems.

For certain problems, a poor initialization can completely kill our computation.
We have encountered this when dealing with likelihoods whose evaluation and differentiation involves a non-trivial numerical operation, such as solving a differential equation.
In some regions of $\theta$'s space, the differential equation may be extremely difficult to solve, requiring a great deal of computation per iteration, even though it is better behaved, with high probability, once we approximately sample from the stationary distribution \citep[e.g.,][]{Margossian:2020b, Margossian:2021}.
This problem manifests as an MCMC run with an extremely slow warmup start before the computation time per iteration improves.
In such a case, a good initialization can bypass pathological regions that frustrate our computation.
For certain problems, values of $\theta$ that pose numerical difficulty are unavoidable, even in the ideal circumstance where we sample from $p$, and the numerical problem cannot be alleviated by any choice of starting distribution.

{\bf Task 3} (diagnostics), in conjunction with diagnostics using multiple chains, encourages a starting distribution with a large variance, also termed an \textit{overdispersed initialization} \citep{Gelman:1992}.
This is a valid strategy for identifying multiple modes and making sure all our Markov chains are not trapped inside the same basin of attraction.
In many cases, a large initial variance also implies the expected squared error of $\bar f$ is dominated by the nonstationary variance, which we can monitor, rather than the unknown squared bias.
This choice of initialization increases the variance of our estimator, but this may be acceptable, since the additional variance decays with the bias and the Markov chains eventually forget their starting point.

A compromise must be struck.
A large variance makes our diagnostics more reliable, however it increases the relaxation time and makes us vulnerable to slow computation in potentially irrelevant regions.
In a Bayesian inference context, we recommend drawing the start of the Markov chain from the prior distribution.
When the data is informative, the posterior tends to be more compact than the prior, and so the prior variance is large relative to the posterior variance.
At the same time, a well-constructed prior often---though not always---excludes the patently absurd values of $\theta$ that can make evaluation and differentiation of a likelihood challenging.
We are not saying that model choices should be determined by computational constraints.
Rather, if a prior allows for values that make a likelihood numerically unstable, it is worth investigating whether we have accounted for all available information when constructing the prior.

\section{Summary and discussion}

Given the challenges of exploring arbitrary distributions in high dimension, it is natural to feel skepticism, demand conservative thresholds for diagnostics, and encourage researchers to continuously increase the length of their Markov chains.
When studying the number of iterations required for MCMC, we find it useful to distinguish three tasks: inference about parameters and quantities of interest ({\bf task 1}), computation of expectation values ({\bf task 2}), and checks to insure the Monte Carlo estimates are accurate ({\bf task 3}).
All three tasks are important, but they can require different amounts of computation.

Crucially, the computation we do not spend on one MCMC run can almost always be put to work elsewhere, for example on a more sophisticated model.
What might be a prudent choice from an inferential perspective can involve a risk from a modeling standpoint.
The field of pharmacometrics offers models with a wide range of sophistication: from the standard model of the human body as a handful of communicating compartments \citep{Gilbaldi:1982} to quantitative systems pharmacology models such as \cite{Peterson:2010}.
In our experience, practical use of Bayesian inference lies somewhere between those two extremes, with so-called semi-mechanistic models taking hours or days to fit with MCMC.
In other words, precise Bayesian inference currently precludes the use of certain models, which we might tackle with cheap and approximate inference, even though this is not ideal.
To take another example, this time from machine learning, there are many compelling reasons to fit a Bayesian neural network \citep{Neal:1996} and the benefits of doing so with MCMC, in terms of accuracy, have been demonstrated \citep{Izmailov:2021}.
But under the constraint of a computational budget, we should wonder whether it is better to fit a small neural network with long MCMC, or a large neural network with a fast approximation.
Ongoing research can, in the longer run, push back against this compromise.

Current MCMC samplers require several conscious decisions from the practitioner, who needs to understand not only their scientific problem but also certain mechanics of MCMC.
A more black box Bayesian inference would leave the user to specify a target ESS, while choices such as the length of the warmup and sampling phases, and even the number of Markov chains are handled automatically under the hood.
This would lead to a completely tuning-free MCMC.
We see many promising directions to implement such a black box MCMC---adaptive warmup and sampling lengths based on online diagnostics \citep{Geweke:1992, Cowles:1998, Jones:2006}, unbiased MCMC using coupling chains \citep{Jacob:2020}, the many-short-chains regime (\cite{Lao:2020}, \cite{Margossian:2023}, and Section~\ref{sec:many-short-chains} of this chapter), and more---and we expect in the future to see some of these ideas implemented in probabilistic programming languages.

\section{Acknowledgments} 

We thank Bob Carpenter, Matt Hoffman, Manny Mokel, Mitzi Morris, Aki Vehtari, and Brian Ward for reading the manuscript and providing feedback, and we thank the U.S. Office of Naval Research for partial support of this work.
The authors are also indebted to the Bayesian computation reading group at the Flatiron Institute and its participants for many fruitful conversations which inspired several passages in this article.

\bibliographystyle{apalike} 
\bibliography{handbook}

\begin{thebibliography}{}

\bibitem[Andrieu and Thoms, 2008]{Andrieu:2008}
Andrieu, C. and Thoms, J. (2008).
\newblock A tutorial on adaptive {MCMC}.
\newblock {\em Statistics and Computing}, 18:343--376.

\bibitem[Badger, 1994]{Badger:1994}
Badger, L. (1994).
\newblock Lazzarini's lucky approximation of $\pi$.
\newblock {\em Mathematics Magazine}, 67:83--91.

\bibitem[Besag and Green, 1993]{Besag:1993}
Besag, J. and Green, P.~J. (1993).
\newblock Spatial statistics and {Bayesian} computation.
\newblock {\em Journal of the Royal Statistical Society, Series B}, 55:25--37.

\bibitem[Betancourt, 2018]{Betancourt:2018}
Betancourt, M. (2018).
\newblock A conceptual introduction to {Hamiltonian Monte Carlo}.
\newblock {\em arXiv:1701.02434v1}.

\bibitem[Blei, 2014]{Blei:2014}
Blei, D.~M. (2014).
\newblock Build, compute, critique, repeat: Data analysis with latent variable
  models.
\newblock {\em Annual Review of Statistics and Its Application}, 1.

\bibitem[Box and Draper, 1987]{Box:1987}
Box, G. E.~P. and Draper, N. (1987).
\newblock {\em Empirical Model-Building and Response Surfaces}.
\newblock Wiley.

\bibitem[Brooks et~al., 2003]{Brooks:2003}
Brooks, S., Giudici, P., and Phillipe, A. (2003).
\newblock Nonparametric convergence assessment for {MCMC} model selection.
\newblock {\em Journal of Computational and Graphical Statistics}, 12:1--22.

\bibitem[Brooks and Gelman, 1998]{Brooks:1998}
Brooks, S.~P. and Gelman, A. (1998).
\newblock General methods for monitoring convergence of iterative simulations.
\newblock {\em Journal of Computational and Graphical Statistics}, 7:434--455.

\bibitem[Campbell et~al., 2023]{outbreaks}
Campbell, F., Frost, S., Jombart, T., Nouvellet, P., Park, S.~W., Pulliam,
  J.~R., Schumacher, J., and Sudre, B. (2023).
\newblock \texttt{outbreaks}: a compilation of disease outbreak data.

\bibitem[Carpenter et~al., 2017]{Carpenter:2017}
Carpenter, B., Gelman, A., Hoffman, M., Lee, D., Goodrich, B., Betancourt, M.,
  Brubaker, M.~A., Guo, J., Li, P., and Riddel, A. (2017).
\newblock Stan: {A} probabilistic programming language.
\newblock {\em Journal of Statistical Software}, 76:1--32.

\bibitem[Cowles et~al., 1998]{Cowles:1998}
Cowles, M.~K., Roberts, G.~O., and Rosenthal, J.~S. (1998).
\newblock Possible biases induced by {MCMC} convergence diagnostics.
\newblock {\em Journal of Statistical Computation and Simulation}, 64:87--104.

\bibitem[du~Ch\'e and Margossian, 2023]{DuChe:2023}
du~Ch\'e, S. and Margossian, C.~C. (2023).
\newblock Parallelization for {Markov} chain {Monte Carlo} with heterogeneous
  runtimes.
\newblock {\em BayesComp, {\tt
  https://charlesm93.github.io/files/Bayescomp\_ode\_chains.pdf}}.

\bibitem[Garbri\'e et~al., 2022]{Gabrie:2022}
Garbri\'e, M., Rotskoff, G.~M., and Vanden-Eijnden, E. (2022).
\newblock Adaptive {Monte Carlo} augmented with normalizing flows.
\newblock {\em Proceedings of the National Academy of Sciences},
  119:e2109420119.

\bibitem[Gelman et~al., 2014]{Gelman:2013}
Gelman, A., Carlin, J.~B., Stern, H.~S., Dunson, D.~B., Vehtari, A., and Rubin,
  D.~B. (2014).
\newblock {\em Bayesian Data Analysis, third edition}.
\newblock CRC Press, London.

\bibitem[Gelman et~al., 2003]{Gelman:2003}
Gelman, A., Carlin, J.~B., Stern, H.~S., and Rubin, D.~B. (2003).
\newblock {\em Bayesian Data Analysis, second edition}.
\newblock CRC Press.

\bibitem[Gelman and Rubin, 1992]{Gelman:1992}
Gelman, A. and Rubin, D.~B. (1992).
\newblock Inference from iterative simulation using multiple sequences (with
  discussion).
\newblock {\em Statistical Science}, 7:457--511.

\bibitem[Gelman and Shirley, 2011]{Gelman:2011}
Gelman, A. and Shirley, K. (2011).
\newblock Inference from simulations and monitoring convergence.
\newblock In Brooks, S., Gelman, A., Jones, G.~L., and Meng, X.-L., editors,
  {\em Handbook of Markov Chain Monte Carlo}, chapter~6. CRC Press.

\bibitem[Gelman et~al., 2020]{Gelman:2020}
Gelman, A., Vehtari, A., Simpson, D., Margossian, C.~C., Carpenter, B., Yao,
  Y., Kennedy, L., Gabry, J., B\"urkner, P.-C., and Modr\'ak, M. (2020).
\newblock Bayesian workflow.
\newblock {\em arXiv:2011.01808}.

\bibitem[Geweke, 1992]{Geweke:1992}
Geweke, J. (1992).
\newblock Evaluating the accuracy of sampling-based approaches to the
  calculation of posterior moments.
\newblock In {\em Bayesian Statistics 4}, pages 169--193. Oxford University
  Press.

\bibitem[Geyer, 1992]{Geyer:1992}
Geyer, C. (1992).
\newblock Practical {Markov} chain {Monte} {Carlo}.
\newblock {\em Statistical Science}, 7:473--483.

\bibitem[Gilbaldi and Perrier, 1982]{Gilbaldi:1982}
Gilbaldi, M. and Perrier, D. (1982).
\newblock {\em Pharmacokinetics, second edition}.
\newblock CRC Press.

\bibitem[Gilks et~al., 1994]{Gilks:1994}
Gilks, W.~R., Roberts, G.~O., and George, E.~I. (1994).
\newblock Adaptive direction sampling.
\newblock {\em Journal of the Royal Statistical Society: Series {D}},
  43:179--189.

\bibitem[Grinsztajn et~al., 2021]{Grinsztajn:2021}
Grinsztajn, L., Semenova, E., Margossian, C.~C., and Riou, J. (2021).
\newblock Bayesian workflow for disease transmission modeling in {Stan}.
\newblock {\em Statistics in Medicine}, 40:6209--6234.

\bibitem[Hoffman and Sountsov, 2022]{Hoffman:2022}
Hoffman, M. and Sountsov, P. (2022).
\newblock Tuning-free generalized {Hamiltonian Monte Carlo}.
\newblock {\em Artificial Intelligence and Statistics}.

\bibitem[Hoffman and Gelman, 2014]{Hoffman:2014}
Hoffman, M.~D. and Gelman, A. (2014).
\newblock The no-{U}-turn sampler: Adaptively setting path lengths in
  {Hamiltonian Monte Carlo}.
\newblock {\em Journal of Machine Learning Research}, 15:1593--1623.

\bibitem[Hoffman et~al., 2021]{Hoffman:2021}
Hoffman, M.~D., Radul, A., and Sountsov, P. (2021).
\newblock An adaptive {MCMC} scheme for setting trajectory lengths in
  {Hamiltonian Monte Carlo}.
\newblock {\em Artificial Intelligence and Statistics}.

\bibitem[Izmailov et~al., 2021]{Izmailov:2021}
Izmailov, P., Vikram, S., Hoffman, M.~D., and Wilson, A.~G. (2021).
\newblock What are {Bayesian} neural network posteriors really like?
\newblock {\em International Conference on Machine Learning}.

\bibitem[Jacob et~al., 2020]{Jacob:2020}
Jacob, P.~E., O'Leary, J., and Atchad\'e, Y.~F. (2020).
\newblock Unbiased {Markov} chain {Monte Carlo} methods with couplings.
\newblock {\em Journal of the Royal Statistical Society, Series B},
  82:543--600.

\bibitem[Jones et~al., 2006]{Jones:2006}
Jones, G.~L., Haran, M., Caffo, B.~S., and Neath, R. (2006).
\newblock Fixed-width output analysis for {Markov} chain {Monte} {Carlo}.
\newblock {\em Journal of the American Statistical Association},
  101:1537--1547.

\bibitem[Lambert and Vehtari, 2022]{Lambert:2022}
Lambert, B. and Vehtari, A. (2022).
\newblock {$R^*$}: A robust {MCMC} convergence diagnostic with uncertainty
  using decision tree classifiers.
\newblock {\em Bayesian Analysis}, 17:353--379.

\bibitem[Landau and Binder, 2009]{Landau:2009}
Landau, D. and Binder, K. (2009).
\newblock {\em A Guide to {Monte Carlo} Simulations in Statistical Physics}.
\newblock Cambridge University Press.

\bibitem[Lao et~al., 2020]{Lao:2020}
Lao, J., Suter, C., Langmore, I., Chimisov, C., Saxena, A., Sountsov, P.,
  Moore, D., Saurous, R.~A., Hoffman, M.~D., and Dillon, J.~V. (2020).
\newblock tfp.mcmc: Modern {Markov} chain {Monte Carlo} tools built for modern
  hardware.
\newblock {\em arXiv:2002.01184}.

\bibitem[Laplace, 1812]{Laplace:1812}
Laplace, P.-S. (1812).
\newblock {\em Th\'eorie Analytique des Probabilit\'es}.

\bibitem[Lee et~al., 2010]{Lee:2010}
Lee, A., Yau, C., Giles, M.~B., Doucet, A., and Holmes, C.~C. (2010).
\newblock On the utility of graphics cards to perform massively parallel
  simulation of advanced {Monte Carlo} methods.
\newblock {\em Journal of Computational and Graphical Statistics}, 19:769--789.

\bibitem[Mackay, 2003]{Mackay:2003}
Mackay, D.~J. (2003).
\newblock {\em Information Theory, Inference, and Learning Algorithms}.
\newblock Cambridge University Press.

\bibitem[Margossian and Gelman, 2020]{Margossian:2020b}
Margossian, C.~C. and Gelman, A. (2020).
\newblock Bayesian model of planetary motion: Exploring ideas for a modeling
  workflow when dealing with ordinary differential equations and multimodality.
\newblock {\em Stan Case Studies}, 7.

\bibitem[Margossian et~al., 2024]{Margossian:2023}
Margossian, C.~C., Hoffman, M.~D., Sountsov, P., Riou-Durand, L., Vehtari, A.,
  and Gelman, A. (2024).
\newblock Nested {$\widehat R$}: Assessing the convergence of {Markov chain
  Monte Carlo} when running many short chains.
\newblock {\em arXiv:2110.13017}.

\bibitem[Margossian et~al., 2020]{Margossian:2020}
Margossian, C.~C., Vehtari, A., Simpson, D., and Agrawal, R. (2020).
\newblock {Hamiltonian Monte Carlo} using an adjoint-differentiated {L}aplace
  approximation: {B}ayesian inference for latent {Gaussian} models and beyond.
\newblock {\em Neural Information Processing Systems}.

\bibitem[Margossian et~al., 2021]{Margossian:2021}
Margossian, C.~C., Zhang, L., Weber, S., and Gelman, A. (2021).
\newblock Solving {ODEs} in a bayesian context: challenges and opportunities.
\newblock {\em Population Approach Group in Europe}.

\bibitem[Mira, 2001]{Mira:2001}
Mira, A. (2001).
\newblock Ordering and improving the performance of {M}onte {C}arlo {M}arkov
  chains.
\newblock {\em Statistical Science}, 16:340--350.

\bibitem[Mira et~al., 2013]{Mira:2013}
Mira, A., Solgi, R., and Imparato, D. (2013).
\newblock Zero-variance principle for monte carlo algorithms.
\newblock {\em Statistics and Computing}, 23.

\bibitem[Moins et~al., 2023]{Moins:2022}
Moins, T., Arbel, J., Dutfoy, A., and Girard, S. (2023).
\newblock On the use of a local {$\widehat R$} to improve {MCMC} convergence
  diagnostic.
\newblock {\em Bayesian Analysis}.

\bibitem[Neal, 1996]{Neal:1996}
Neal, R.~M. (1996).
\newblock {\em Bayesian Learning for Neural Networks}.
\newblock Springer.

\bibitem[Neal, 2011]{Neal:2011}
Neal, R.~M. (2011).
\newblock {MCMC} using {H}amiltonian dynamics.
\newblock In {\em Handbook of Markov Chain Monte Carlo}. CRC Press.

\bibitem[Pasarica and Gelman, 2010]{Pasarica:2010}
Pasarica, C. and Gelman, A. (2010).
\newblock Adaptively scaling the {Metropolis} algorithm using expected squared
  jumped distance.
\newblock {\em Statistica Sinica}, 20:343–364.

\bibitem[Peterson and Riggs, 2010]{Peterson:2010}
Peterson, M.~C. and Riggs, M.~M. (2010).
\newblock A physiologically based mathematical model of integrated calcium
  homeostasis and bone remodeling.
\newblock {\em Bone}, 46:49--63.

\bibitem[Piironen and Vehtari, 2017]{Piironen:2017}
Piironen, J. and Vehtari, A. (2017).
\newblock Sparsity information and regularization in the horseshoe and other
  shrinkage priors.
\newblock {\em Electronic Journal of Statistics}, 11:5018--5051.

\bibitem[Riou-Durand et~al., 2023]{Riou-Durand:2023}
Riou-Durand, L., Sountsov, P., Vogrinc, J., Margossian, C.~C., and Power, S.
  (2023).
\newblock Adaptive tuning for {Metropolis} adjusted {Langevin} trajectories.
\newblock {\em Artificial Intelligence and Statistics}.

\bibitem[Roberts et~al., 1997]{Gelman:1997}
Roberts, G.~O., Gelman, A., and Gilks, W.~R. (1997).
\newblock Weak convergence and optimal scaling of random walk {Metropolis}
  algorithms.
\newblock {\em Annals of Applied Probability}, 7:110--120.

\bibitem[Roberts and Rosenthal, 1998]{Roberts:1998}
Roberts, G.~O. and Rosenthal, J.~S. (1998).
\newblock Optimal scaling of discrete approximations to {Langevin} diffusions.
\newblock {\em Journal of the Royal Statistical Society, Series B},
  60:255--268.

\bibitem[Roberts and Rosenthal, 2004]{Roberts:2004}
Roberts, G.~O. and Rosenthal, J.~S. (2004).
\newblock General state space {Markov chains and MCMC} algorithms.
\newblock {\em Probability Surveys}, 1:20--71.

\bibitem[Rosenthal, 2000]{Rosenthal:2000}
Rosenthal, J.~S. (2000).
\newblock Parallel computing and {Monte Carlo} algorithms.
\newblock {\em Far East Journal of Theoretical Statistics}, 4:207--236.

\bibitem[Salvatier et~al., 2016]{Salvatier:2016}
Salvatier, J., Wiecki, T.~V., and Fonnesbeck, C. (2016).
\newblock Probabilistic programming in {Python} using {PyMC3}.
\newblock {\em PeerJ Computer Science}, 2.

\bibitem[Sountsov and Hoffman, 2021]{Sountsov:2021}
Sountsov, P. and Hoffman, M.~D. (2021).
\newblock Focusing on difficult directions for learning {HMC} trajectory
  lengths.
\newblock {\em arXiv:2110.11576}.

\bibitem[South et~al., 2023]{South:2023}
South, L.~F., Oates, C.~J., Mira, A., and Drovandi, C. (2023).
\newblock Regularized zero-variance control variates.
\newblock {\em Bayesian Analysis}, 18:865--888.

\bibitem[South et~al., 2021]{South:2021}
South, L.~F., Riabiz, M., Teymur, O., and Oates, C.~J. (2021).
\newblock Post-processing of {MCMC}.
\newblock {\em Annual Review of Statistics and Its Application}, 9:1--30.

\bibitem[{TensorFlow Probability Development Team}, 2023]{tfp:2023}
{TensorFlow Probability Development Team} (2023).
\newblock Tensorflow probability.

\bibitem[Vats et~al., 2019]{Vats:2019}
Vats, D., Flegal, J.~M., and Jones, G.~L. (2019).
\newblock Multivariate output analysis for {Markov chain Monte Carlo}.
\newblock {\em Biometrika}, 106:321--337.

\bibitem[Vats and Knudson, 2021]{Vats:2021}
Vats, D. and Knudson, D. (2021).
\newblock Revisiting the {Gelman-Rubin} diagnostic.
\newblock {\em Statistical Science}, 36:518--529.

\bibitem[\v{C}e\v{s}novar et~al., 2020]{Cesnovar:2020}
\v{C}e\v{s}novar, R., Bronder, S., Sluga, D., Dem\v{s}ar, Ciglari\v{c}, T.,
  Talts, S., and \v{S}trumbelj, E. (2020).
\newblock {GPU}-based parallel computation support for {Stan}.
\newblock {\em arXiv:1907.01063v2}.

\bibitem[Vehtari, 2022]{Vehtari:2022}
Vehtari, A. (2022).
\newblock Bayesian workflow book - digits.

\bibitem[Vehtari et~al., 2021]{Vehtari:2021}
Vehtari, A., Gelman, A., Simpson, D., Carpenter, B., and B\"urkner, P.-C.
  (2021).
\newblock Rank-normalization, folding, and localization: An improved {$\widehat
  R$} for assessing convergence of {MCMC} (with discussion).
\newblock {\em Bayesian Analysis}, 16:667--718.

\bibitem[\v{S}trumbelj et~al., 2023]{Strumbelj:2023}
\v{S}trumbelj, E., Bouchard-C\^{o}t\'{e}, A., Corander, J., Gelman, A.,
  H\.{a}vard, R., Murray, L., Pesonen, H., Plummer, M., and Vehtari, A. (2023).
\newblock Past, present, and future of software for {Bayesian} inference.
\newblock {\em Statistical Science}.

\bibitem[Zhang et~al., 2022]{Zhang:2022}
Zhang, L., Carpenter, B., Gelman, A., and Vehtari, A. (2022).
\newblock Pathfinder: Parallel quasi-{N}ewton variational inference.
\newblock {\em Journal of Machine Learning Research}, 23(306):1--49.

\end{thebibliography}

\end{document}